\documentclass[twocolumn]{aastex631}

\usepackage{amsmath}
\usepackage{hyperref}
\usepackage{CJKutf8}

\usepackage[normalem]{ulem}

\newcommand{\caltech}{Department of Astronomy, California Institute of Technology, Pasadena, CA 91125, USA}

\newcommand{\ucsd}{Department of Astronomy \& Astrophysics,  University of California, San Diego, La Jolla, CA 92093, USA}
\newcommand{\nmsu}{Department of Astronomy, New Mexico State University, PO Box 30001, MSC 4500, Las Cruces, NM 88003, USA}

\begin{document}

\title{On the Orbit of the Binary Brown Dwarf Companion GL229 Ba and Bb}

\author[0000-0001-5684-4593]{William Thompson}
\affiliation{NRC Herzberg Astronomy and Astrophysics,
5071 West Saanich Road,
Victoria, BC, V9E 2E7, Canada}

\author[0000-0001-9582-4261]{Dori Blakely}
\affiliation{Department of Physics and Astronomy, University of Victoria,
3800 Finnerty Road, Elliot Building,
Victoria, BC, V8P 5C2, Canada}
\affiliation{NRC Herzberg Astronomy and Astrophysics,
5071 West Saanich Road,
Victoria, BC, V9E 2E7, Canada}

\author[0000-0002-6618-1137]{Jerry W. Xuan}
\affiliation{\caltech}

\author[0000-0001-7402-8506]{Alexandre Bouchard-Côté}
\affiliation{Department of Statistics, University of British Columbia, 2329 West Mall, Vancouver BC, Canada, V6T 1Z4.}

\author{Guillaume Bourdarot}
\affiliation{Max Planck Institute for extraterrestrial Physics,
Giessenbachstrasse, 85748 Garching, Germany}

\author[0000-0003-1001-1216]{Miguel Biron-Lattes}
\affiliation{Department of Statistics, University of British Columbia, 2329 West Mall, Vancouver BC, Canada, V6T 1Z4.}

\author[0000-0003-1499-0191]{Trevor Campbell}
\affiliation{Department of Statistics, University of British Columbia, 2329 West Mall, Vancouver BC, Canada, V6T 1Z4.}

\author{Frank Eisenhauer}
\affiliation{Max Planck Institute for extraterrestrial Physics,
Giessenbachstrasse, 85748 Garching, Germany}

\author[0000-0002-1493-300X]{Thomas Henning}
\affiliation{MPI for Astronomy, K\"onigstuhl 17, 69117 Heidelberg, Germany}

\author[0000-0001-8345-593X]{Markus Janson}
\affiliation{Department of Astronomy, Stockholm University, AlbaNova University Center, 10691 Stockholm, Sweden}

\author[0000-0002-6773-459X]{Doug Johnstone}
\affiliation{NRC Herzberg Astronomy and Astrophysics,
5071 West Saanich Road,
Victoria, BC, V9E 2E7, Canada}
\affiliation{Department of Physics and Astronomy, University of Victoria,
3800 Finnerty Road, Elliot Building,
Victoria, BC, V8P 5C2, Canada}

\author[0000-0003-2769-0438]{Jens Kammerer}
\affiliation{European Southern Observatory, Karl-Schwarzschild-Str. 2, 85748 Garching, Germany}

\author[0000-0002-9936-6285]{Quinn Konopacky}
\affiliation{\ucsd}

\author[0000-0002-6948-0263]{Sylvestre Lacour}
\affiliation{LESIA, Observatoire de Paris, PSL, CNRS, Sorbonne Universit\'e, Universit\'e de Paris, 5 place Janssen, 92195 Meudon, France}

\author[0000-0002-4164-4182]{Christian Marois}
\affiliation{NRC Herzberg Astronomy and Astrophysics,
5071 West Saanich Road,
Victoria, BC, V9E 2E7, Canada}
\affiliation{Department of Physics and Astronomy, University of Victoria,
3800 Finnerty Road, Elliot Building,
Victoria, BC, V8P 5C2, Canada}

\author[0000-0002-8895-4735]{Dimitri Mawet}
\affiliation{\caltech}

\author[0000-0003-2125-0183]{Antoine Mérand}
\affiliation{European Southern Observatory, Karl-Schwarzschild-Str. 2, 85748 Garching, Germany}

\author[0000-0002-9242-9052]{Jayke Samson Nguyen}
\affiliation{\ucsd}

\author{Eric Nielsen}
\affiliation{\nmsu}

\author[0000-0003-4203-9715]{Emily Rickman}
\affiliation{European Space Agency (ESA), ESA Office, Space Telescope Science Institute, 3700 San Martin Drive, Baltimore, MD 21218, USA}

\author[0000-0003-2233-4821]{Jean-Baptiste Ruffio}
\affiliation{\ucsd}

\author[0000-0002-5629-7910]{Nikola Surjanovic}
\affiliation{Department of Statistics, University of British Columbia, 2329 West Mall, Vancouver BC, Canada, V6T 1Z4.}

\author[0000-0003-0774-6502]{Jason J. Wang
\begin{CJK*}{UTF8}{gbsn}(王劲飞)\end{CJK*}}
\affil{Center for Interdisciplinary Exploration and Research in Astrophysics (CIERA), Northwestern University,
1800 Sherman Ave, Evanston, IL, 60201, USA}
\affil{Department of Physics and Astronomy, Northwestern University, 2145 Sheridan Rd, Evanston, IL 60208, USA}

\author[0000-0002-9764-5109]{Thomas Winterhalder}
\affiliation{European Southern Observatory, Karl-Schwarzschild-Str. 2, 85748 Garching, Germany}

\begin{abstract}
The companion GL229 B was recently resolved by \citet{xuannat} as a tight binary of two brown dwarfs (Ba and Bb) through VLTI-GRAVITY interferometry and VLT-CRIRES+ radial velocity measurements. Here, we present Bayesian models of the interferometric and radial velocity data in additional detail, along with an updated outer orbit of the brown dwarf pair about the primary. To create a Bayesian model of the inner orbit with appropriate uncertainties, we develop an application of kernel phases to GRAVITY data to address baseline redundancy in closure phase modelling. By using a modern parallel tempered MCMC algorithm, we show that we can  constrain the binary's orbit using only the full complement of GRAVITY data, despite each epoch having insufficient visibility-plane coverage and/or SNR to determine a unique position. Based on these methods, we demonstrate excellent agreement between independent analyses of the GRAVITY and CRIRES+ datasets, which mutually resolve key parameter degeneracies to yield a single well-constrained orbital solution.
The inner binary has a period of 12.1346±0.0011 days, eccentricity of 0.2317±0.0025, and total mass of 71.0±0.4 $\mathrm{M_{Jup}}$, with Ba and Bb having masses of 37.7±1.1 $\mathrm{M_{Jup}}$ and 33.4±1.0 $\mathrm{M_{Jup}}$ respectively. We also present an updated model of GL229 B's outer orbit around the primary M-dwarf A, incorporating new Keck/NIRC2 astrometry. We find a semi-major axis of $42.9^{+3.0}_{-2.4}$ AU, eccentricity of 0.736±0.014, and a total mass for B of ($71.7\pm0.6 \mathrm{M_{Jup}}$) that is consistent with that derived from the inner orbit. 
We calculate a mutual inclination of $31±2.5^\circ$ between the inner and outer orbits, below the threshold for Kozai-Lidov oscillations.
The agreement on the mass of Ba + Bb between the inner and outer orbits is an important test of our ability to model RV, astrometry, and Hipparcos-Gaia proper motion anomaly.
Our methodological advances, particularly in handling low SNR and sparse UV-coverage interferometric data, will benefit future observations of rapidly-orbiting companions with GRAVITY.
\end{abstract}
s
\section*{}
\clearpage
\section{Introduction} \label{sec:intro}
GL229 is a hierarchical triple system, consisting of a central M-dwarf A and a companion B \citep{gl229b_discovery_Nakajima,gl229b_discovery_Oppenheimer} which was recently resolved as tight binary of two brown dwarfs, Ba and Bb, orbiting just 0.04 AU apart contemporaneously by \citet{xuannat} and by \citet{Whitebook_2024}.
This detection resolves a historical tension between the luminosity and dynamical mass of the companion, noted by \citet{tdbrandt_2020} and investigated by \citet{howe_gl229_binary}.

\citet{xuannat} determined that the B companion was a binary through unresolved radial velocity measurements and through direct detection of the binary components by VLTI-GRAVITY. 
The modelling of this data proved challenging for two reasons.
First, the radial velocity data from CRIRES+ are sampled relatively sparsely in time, consisting of only seven points for both objects.
Second, the GRAVITY data available possesses a relatively low signal to noise ratio, has limited coverage in the visibility plane at most epochs, and does not include calibrated visibility amplitudes, only closure phases.
As a result, the GRAVITY data is consistent with many different locations at each epoch.

\citet{xuannat} presented joint models of the RV and GRAVITY data using both a frequentist approach using bootstrapping, and a fully Bayesian model. 
Here we present Bayesian models of GL229 B in substantially more detail.

The sparsity and multi-modality of the data raises questions as to whether we can robustly identify the correct orbit and dynamical mass when each individual GRAVITY epoch is consistent with many companion locations. 
In this work, we will argue that we can.

To start, we motivate and present an application of interferometric kernel phases to GRAVITY, showing how these allow us to address a common problem with closure phase modelling caused by baseline redundancy \citep[noted in the GRAVITY context by ][]{jens}.
Next, we describe our use of an improved MCMC algorithm and software package that allows us to globally integrate the multi-modal GRAVITY posterior.

We then present our findings. These include independent analyses of the RV and of the GRAVITY data to demonstrate their excellent agreement, a discussion on how flux modelling choices impact the inferred orbital parameters, and an updated outer orbit of GL229 B around the primary A.
We then show that the new independent constraint on the total dynamical mass of B is consistent with the mass derived from the outer orbit through astrometry; HARPS, HIRES, Lick, and UVES RVs; and Hipparcos-Gaia proper motion anomaly.
Finally, we present an updated calculation of the mutual inclination between the inner orbit of GL229 Ba Bb and their outer orbit around GL229 A.

The results presented herein build confidence in the strongly constrained orbit and dynamical mass measurements of Gl229 Ba and Bb, and the updated outer orbit and mutual inclination may help constrain the dynamical history of the system.

The methodology we present is widely applicable to future interferometric measurements of planets and stars with low SNR and/or low visibility plane coverage.
This is especially important for systems which have rapid orbital motion which will be increasingly common as GRAVITY continues to probe fainter and closer-in planets.
All code is made publicly available through a new release of Octofitter \citep{octofitter}\footnote{\href{https://sefffal.github.io/Octofitter.jl/}{https://sefffal.github.io/Octofitter.jl/}}, version 4.

\section{Observations}

The observations used in the analysis of the inner binary are the same as in \citet{xuannat}, namely five epochs of GRAVITY observations and seven epochs of CRIRES+ radial velocities. The data are available at Zenodo\footnote{\href{https://zenodo.org/records/13851639}{https://zenodo.org/records/13851639}} \citep{xuan_2024_zenodo} .

\begin{table}
\caption{Epochs and times of GRAVITY observations from \citet{xuannat}.\label{tab:obs-grav}}
    \centering
    \begin{tabular}{|l|}
        \hline
        date (UTC) \\
        \hline
        First epoch \\
        \hline
        2023-12-26 04:13:56.643 \\
        2023-12-26 04:31:08.687 \\
        2023-12-26 04:46:32.726 \\
        2023-12-26 04:53:47.744 \\
        2023-12-26 05:00:59.763 \\
        2023-12-26 05:15:41.800 \\
        2023-12-26 05:36:14.852 \\
        2023-12-26 05:58:44.909 \\
        2023-12-26 06:19:08.961 \\
        2023-12-26 06:27:05.981 \\
        2023-12-26 06:34:23.999 \\
        2023-12-26 06:41:39.018 \\
    
        \hline
        Second epoch \\
        \hline
        2023-12-30 05:47:14.016 \\
        2023-12-30 06:11:23.077 \\
        2023-12-30 06:18:38.095 \\
        2023-12-30 06:25:53.113 \\
    
        \hline
        Third epoch \\
        \hline
        2024-02-28 02:57:28.592 \\
        2024-02-28 03:13:28.632\\
        2024-02-28 03:29:49.673\\
    
        \hline
        Fourth epoch \\
        \hline
        2024-03-29 00:31:13.894 \\
    
        \hline
        Fifth epoch \\
        \hline
        2024-04-29 23:15:12.061 \\
        2024-04-29 23:22:24.079 \\
        \hline
    \end{tabular}
\end{table}

We conducted the GRAVITY observations under the program IDs 0112.C-2369(A) and 2112.D-5036(A) (PI: Xuan).
The data were reduced using the ESO GRAVITY pipeline v1.6.4, as described in \cite{xuannat}.
See Table \ref{tab:obs-grav} for a full listing (reproduction of \citet{xuannat}, extended data table 1).
In dual-field wide mode, light from the primary GL229A was used for fringe-tracking while a second fiber, placed over the photocentre of GL229B, was directed into the GRAVITY spectrograph. 
In this configuration, the squared visibilities are not well-calibrated, but thankfully closure phases can still be used \citep[see][]{ireland2013}.

Within K-band, the emission spectra of T dwarfs like these objects have a peak near 2.1 $\mathrm{\mu m}$ due to $\mathrm{H_2 O}$ and $\mathrm{CH_4}$ absorption bands to the blue and red respectively. Accordingly, the closure phase data had the highest SNR between $2.025 \;\mathrm{\mu m} < \lambda < 2.15  \;\mathrm{\mu m}$. The data outside this range exhibited high closure phase noise, so we discarded it from our analysis.

The radial velocity values are extracted from CRIRES+ spectra as described below.
Table \ref{tab:rvdat} presents the extracted RVs for both components (reproducing Extended Data Table 3 from \citet{xuannat}).

\begin{table*}[]
\caption{CRIRES+ RVs reproduced from \citet{xuannat}.\label{tab:rvdat}}
    \centering
    \begin{tabular}{|l|r|r|r|r|}
    \hline
    date (UTC)  & $v_{B_a}$  (km/s)& $v_{B_b}$ (km/s) & $\sigma_{v_{B_a}}$ (km/s) & $\sigma_{v_{B_b}}$ (km/s)  \\
    \hline
    2024-02-19 03:15:47    &     8.07  &  -7.51  &   0.12  &  0.19 \\
    2024-02-20 02:37:32    &     8.13  &  -7.22  &   0.14  &  0.19 \\
    2024-03-01 02:33:20    &     6.07  &  -6.31  &   0.17  &  0.25 \\
    2024-03-19 01:42:02    &    -3.21  &   4.09  &   0.54  &  0.75  \\
    2024-03-20 01:23:10    &    -8.97  &  10.60  &   0.23  &  0.46  \\
    2024-04-07 00:45:32    &     6.85  &  -7.55  &   0.16  &  0.24  \\
    2024-04-07 23:58:08    &     7.49  &  -8.10  &   0.27  &  0.32  \\
    \hline
    \end{tabular}
\end{table*}

Regarding the outer orbit, we further captured a short sequence of K band images using Keck/NIRC2 on 2024-10-24 for the purposes of extracting relative astrometry. We captured both long exposure images to reveal the companion B while saturating the primary A and short exposure images to reveal the primary.

\section{Methods}

\subsection{GRAVITY Modelling and Application of Kernel Phases}

We modelled the complex visibilities from the GRAVITY data using closure phases.

Closure phases are formed by summing the phases along closed triangles, each consisting of three baselines.
By this construction, station-based phase gain is cancelled out making closure phases a more robust observable.
For VLTI, there are four UTs (Unit Telescopes) giving a total of 6 baselines, and four closure phases.
Closure phase modelling of GRAVITY data has been implemented into several software packages, including fouriever \citep[][single-epoch]{fouriever} and PMOIRED \citep[][multi-epoch]{pmoired}.

For VLTI operating with the four UTs, the ``design matrix'' $T$ that specifies how to construct closure phases is given by
\begin{equation}
T = 
\begin{pmatrix}
    1 & -1 &  0 &  1 &  0 & 0\\
    1 & 0  & -1 &  0 &  1 & 0\\
    0 & 1  & -1 &  0 &  0 & 1\\
    0 & 0  & 0  & 1  & -1 & 1\\
\end{pmatrix}
\end{equation}
\noindent This matrix shows how to add or subtract the phases along each of the six baseline to compute the four possible closure phases.

When working with closure phases from GRAVITY, there are two important sources of correlation noted by \citet{jens} that should be considered.
The first is spectral correlation. 
\citet{jens} find a correlation of textbf{0.017} between wavelengths for the same closure phase triangle.
The second is a $\pm1/3$ correlation between closure phases that share a baseline.

This baseline correlation poses a challenge: because they share baselines, the full set of constructable closure phases is larger than the number of independent degrees of freedom in the data.
This presents both a practical challenge (the covariance matrix is degenerate and cannot properly define a multivariate normal distribution) and a statistical challenge (the correlation and redundancy lead to overconfidence and underestimated uncertainties).

One approach that has been proposed as a way to mitigate these correlations is to use  bootstrapping \citep{bootstrap,pmoired}.
Here, the underlying data points are sampled with replacement, and a maximum-likelihood model is locally re-optimized after each draw.
After thousands of repetitions, the fit parameters asymptotically approach the sampling distribution given various assumptions. It has been proposed that this procedure may also account for correlation between closure phase measurements, though it is possible the inflated variance introduced by the resampling may account for the broader confidence intervals seen by previous authors.

Here, we instead address correlation and redundancy directly by reformulating the problem into a set of non-redundant observables: the kernel phases \citep{kernphase_martinache}.
From any given set of closure phases, a kernel phase basis can be defined by identifying a linearly independent subspace.
The transformation from closure phases to kernel phases can be found by several methods, including direct construction of a non-redundant set and through matrix factorization \citep[for an excellent discussion, see ][]{closurestatistics_blackburn}.

The most numerically expedient kernel phase basis is typically found by factorizing the closure phase covariance matrix and discarding singular entries \citep[see][]{laugier_kp_white}. This finds a KP basis where each KP is statistically independent (i.e. the covariance matrix is diagonal in that basis), giving simpler and faster code.
In our case, we would like to derive the covariance matrix from the data so instead we choose our kernel phase basis by Choleksy factorizing the closure phase design matrix and removing singular values. This gives 3 linearly-independent kernel phases per wavelength.
The transformation matrix $P_1$ we find that maps from the four closure phases (specified by $T$ above) to the three independent kernel phases is:
\begin{equation}
P_1 = 
\begin{pmatrix}
 1   & 0.333333  & -0.333333 &   0.333333 \\
 0   & 0.942809  &  0.471405 &  -0.471405 \\
 0   & 0.0       &  0.816497 &   0.816497 \\
\end{pmatrix}
\end{equation}

This can be multiplied through with T to create a new pseudo-``design matrix'' that could be substituted into existing codes:
\begin{multline}
T_{KP} = P_1 T = \\
\begin{pmatrix}
 1.333 & -1.333  &  0         &  1.333  &  0         &  0  \\
 0.942 &  0.471 & -1.414  & -0.471 &  1.414  &  0.0 \\
 0         &  0.816 & -0.816 &  0.816 & -0.816 &  1.633 \\
\end{pmatrix}
\end{multline}

The first row is identifiable as the first closure phase simply scaled by a factor of 1 + 1/3. The two subsequent rows involve less obvious combinations of phases from the six different baselines.
Since the design matrix is the same for all GRAVITY observations, this kernel phase basis is stable and could be used for different programs.

\begin{figure}
    \centering
    \includegraphics[width=\linewidth]{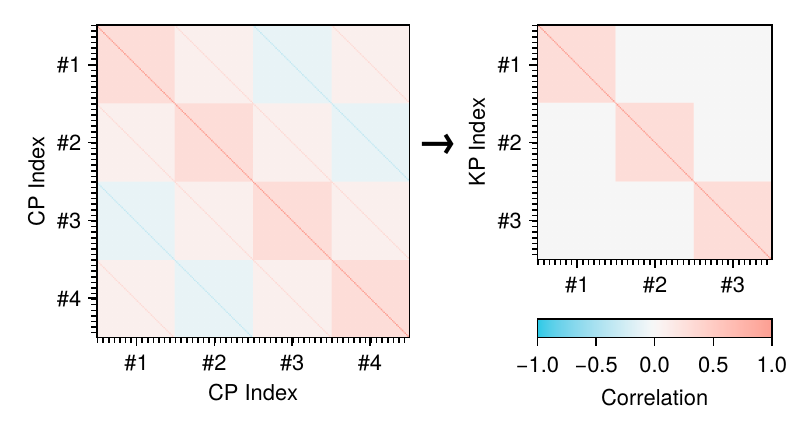}
    \caption{The semi-analytic correlation matrix from \citet{jens} adopted in our model for closure phases (left) and projected into kernel phases (right). The indices run first over wavelength, and then by closure or kernel phase index.
    The closure phase correlation matrix on the left is rank deficient and therefore cannot be used to define a multivariate normal distribution for MCMC sampling. Projecting it onto a non-redundant kernel phase basis solves this issue.}
    \label{fig:corr-mat}
\end{figure}

This kernel phase basis is constructed to be statistically independent for each wavelength, but correlation between different wavelengths for the same kernel phase remain.
We have an insufficient number of observations to estimate a full-rank covariance matrix empirically, and instead model the remaining spectral correlation between kernel phases using an analogue of the semi-analytic matrix presented by \citet{jens}. 
We create a block diagonal correlation matrix parameterized by a kernel phase correlation parameter $C_z$, representing the correlation between wavelengths for the same kernel phase index. We make the $C_z$ a parameter in our model with a uniform prior from 0 to 1.

Figure \ref{fig:corr-mat} shows the correlation matrix defined by \citet{jens} and the same matrix projected into our chosen kernel phase basis through multiplication by $P_1$. This neatly illustrates how adopting kernel phases solves the redundancy issue with closure phase modelling. 

For a given level of closure phase spectral correlation $C_y$ there also exists a kernel phase basis that would have only a diagonal structure, that is, would have each index be statistically independent. We do not adopt that basis since we allow the level of spectral correlation to vary in our model.

We also add a diagonal term to the covariance matrix we call ``kernel phase jitter'' that is added in quadrature to the kernel phase uncertainties, separate for each epoch.
The kernel phase jitter is intended to absorb systematic errors caused by any remaining mis-calibration of the kernel phase data.

In order to compute the likelihood, the covariance matrix must be factorized. Our covariance matrix is parameterized by the kernel phase spectral correlation $C_z$ and the kernel phase jitter, so this factorization must be computed at each likelihood evaluation.
Through profiling, we determined that this step dominates the runtime of our model.
To speed this step up, we exploited the block-diagonal structure of the covariance matrix.
We factorized the three non-zero blocks independently and merged the result together.
This significantly improved the model evaluation speed, but we note that it still remains one of the most expensive steps in the model evaluation.
Since the typical spectral correlation is on the order of $2\times10^{-2}$ \citep{jens}, many practical applications might choose to ignore the spectral correlation and approximate the covariance matrix as diagonal.
Future work could also investigate sampling factors of the covariance matrix directly from a Cholesky-valued distribution \citep[e.g.][]{lkj_dist}.

Lastly, we account for the differential fiber coupling efficiency between the two objects using the theoretical curve for a single-mode fiber, and assuming the fiber is placed at the photocentre at each epoch \citep{jason_pds70_grav}.

\subsection{Radial Velocity Modelling of Binary}
 
The radial velocities of Ba and Bb are extracted from CRIRES+ spectra as described in \citet{xuannat}. 
The binary is spatially unresolved by this instrument, so the radial velocities of both components were spectrally resolved by jointly fitting a model of two brown dwarfs plus contamination from the primary A. 
A table of extracted RVs for both components is available in machine readable form alongside \citet{xuannat}.

We model these two sets of RV measurements in Octofitter as a combination of a barycentric RV for Ba and a relative RV between Bb and Ba. We fit an instrumental RV offset term for the barycentric RV, and an empirical RV jitter term (shared for both sets of observations).

We do not include the effects of the outer orbit around the primary A since we estimate the change in RV to the B barycenter is only 6.3 m/s over the timescale of the CRIRES+ observations, which is well below the measurement uncertainties.

The \citet{Whitebook_2024} RV measurements of B assumed that Ba dominated the flux of Bb, and so fit their spectra using a single template spectrum. Since we now know that both Ba and Bb contribute significant flux, we do not include their measurments in our analysis.

\newpage
\subsection{Binary Model Descriptions}

We implemented both the kernel phase and RV models as likelihood objects in the Octofitter framework, and tied them together into joint models.
In order to examine how the different datasets and our modelling choices impact the resulting posteriors, we consider two suites of models.

The first suite of models aims to check that the CRIRES+ RV and GRAVITY kernel phase data are consistent with each other. We fit a model to the CRIRES+ RV data (``RV only''), to the GRAVITY kernel phase data (``KP only''), and to both data sources (``RV \& KP''). For this comparison, we made simplifying assumptions that the contrast between Ba and Bb was the same across wavelengths and epochs, and that the correlation between wavelength channels within a given kernel phase can be neglected. The results from the ``RV \& KP'' model are exactly those that were shared in the discovery paper \citep{xuannat}.

Some parameters, like contrast, mass ratio, and position angle of ascending node, are only constrained by one of the two data sources and thus will simply follow the prior distribution for the RV only and KP only models respectively.
On the other hand, some parameters like total mass, orbital period, eccentricity, and argument of periastron are independently constrained by both datasets and we should expect to find agreement between the two models.
 
The second suite of models include both the RV and GRAVITY datasets, but then consider how more detailed modelling choices of the GRAVITY data impact the resulting parameter distribution. Thus, where in the first suite of models the contrast between Bb and Ba was assumed to be the same across all wavelengths and exposures, this assumption is relaxed in the second suite so that we may examine its impact on the posterior, and furthermore, the correlation between spectral channels is treated as a parameter that can take on the best value from the data.

In this second suite, we consider three models.
The first model, ``contrast steady,'' assumes that contrast between Bb and Ba is consistent over both wavelength and epoch but now allows the correlation between wavelength channels to be non-zero.
The second model, ``contrast flexible vs. time,'' allows the contrast to vary independently between the five GRAVITY epochs (but not between each individual exposure).
The third model, ``contrast flexible vs. wavelength,'' allows the contrast to vary independently between spectral channels but is otherwise fixed over time.
A fourth plausible model to consider would be allowing the contrast to vary over both time and wavelength, but this is not computationally feasible with our current modelling tools. It is also unclear if such a highly parameterized model is of scientific interest. 

Unlike the first suite, disagreement between these models would offer less obvious interpretation.
Variation in contrast between epochs could be caused either by uncalibrated systematic errors in our data, large fiber positioning errors, or true astrophysical flux variations.
Brown dwarfs can be variable over timescales of hours and longer due to modulation of their flux due to rotation and variability amplitudes of up to approximately 25\% have been observed in J-band \citep[e.g. ][]{radigan2012}. That said, late T dwarfs are rarely variable sources, with maximum variability amplitudes of less than 5\% in the J and H bands (2M2228: \citet{Buenzli2012}, Ross 458 c: \citet{Manjavacas2019}).
Accordingly, if significant variations in contrast over time are detected, we should first suspect systematics.
Some amount of contrast variation versus wavelength is expected and allowing for this variation in our model might allow the orbit parameters to take on different values. 

Beyond these two suites of models, we consider a seventh and final model of the inner Ba/Bb binary.
The preceding six models all adopt a prior on the total mass of the GL229B companion from \citet{gmbrandt} of $71.4 \pm 0.6 \;\mathrm{M_{jup}}$, based on relative astrometry, RV, and proper motion anomaly of the primary A.
In order to assess the impact of this prior, we fit a model with a uniform prior on the total mass of $50..150 \;\mathrm{M_{jup}}$. 
We once more treat the contrast as consistent across wavelengths and epochs, and allow the kernel phase spectral correlation parameter $C_z$ to be non-zero.

The parameters and priors adopted by all of these models are listed in Table \ref{tab:priors}.
We adopt the $\tau$ parameterization first described in \citet{blunt_orbitize_2020}, which replaces $t_0$, the epoch of periastron passage, with a unitless value defined between 0-1, with 1 wrapping back to zero, representing the fraction of the orbital period completed on a specified reference epoch. This value is well-defined even for orbits that have been observed over only a small fraction of their period (like GL229B), since it de-correlates semi-major axis from the current location of the planet.

\begin{table*}
    \caption{Model variables and prior distributions. The parameters listed under the heading ``Repeated in all models'' are defined the same way for all models. The following sub-sections specify the priors on variables that differ between models}.\label{tab:priors}
    \begin{tabular}{|l|l|l|}
        \hline
        \textbf{parameter} & \textbf{description} & \textbf{distribution} \\
        \hline
        \textbf{Repeated in all models} &&\\
        $q$ & mass textbf{ratio} of Bb over Ba & Normal(0.5, 0.5), truncated to [0,1]\\
        $\bar{\omega}$ & parallax [mas]& Normal(173.5740, 0.0170)\textsuperscript{*} \\
        $e$ & eccentricity & Uniform(0,0.999)\\
        $P$ & period [days] & Uniform(5, 20) \\
        $i$ & inclination [rad] & Sine \\
        $\omega$  & argument of periapsis of Bb [rad] & Uniform(0,$2\pi$)\\
        $\Omega$  & position angle of ascending node [rad]& Uniform(0,$2\pi$) \\
        $\tau$  & orbit fraction at MJD 60304.17635003472 & Uniform(0,1)\\
        $\rm{jitter}$  & R.V. jitter [m/s] & LogUniform(1, 5000) \\
        $\rm{rv}_0$  & R.V. offset [m/s] & Normal(0, 2000)\\
        $C_z$ & Kernel phase spectral correlation & Uniform(0,1)\\
        $KP_{jit,1}$ & Kernel phase jitter, epoch 1& Uniform(0, 180) \\
        $KP_{jit,2}$ & Kernel phase jitter, epoch 2& Uniform(0, 180) \\
        $KP_{jit,3}$ & Kernel phase jitter, epoch 3& Uniform(0, 180) \\
        $KP_{jit,4}$ & Kernel phase jitter, epoch 4& Uniform(0, 180) \\
        $KP_{jit,5}$ & Kernel phase jitter, epoch 5& Uniform(0, 180) \\
        \hline
        \textbf{Basic models (RV, KP, or RV \& KP)} & &\\
        $M$ & total mass [$M_{jup}$] & Normal(71.4, 0.6)\textsuperscript{\textdagger} \\
        $f$  & contrast between Bb and Ba & Uniform(0,1) \\
        $C_z$  & Kernel phase spectral correlation & 0 \\
        \hline
        \textbf{Consistent contrast model} & &\\
        $M$ & total mass [$M_{jup}$] & Normal(71.4, 0.6)\textsuperscript{\textdagger} \\
        $f$  & contrast between Bb and Ba & Uniform(0,1) \\
        $C_z$  & Kernel phase spectral correlation & Uniform(0, 1) \\
        \hline
        \textbf{Contrast per epoch model} & &\\
        $M$ & total mass [$M_{jup}$] & Normal(71.4, 0.6)\textsuperscript{\textdagger} \\
        $C_z$  & Spectral correlation parameter & Uniform(0, 1) \\
        $f_i, i \in 1..N_{\rm{epochs}}$  & contrast between Bb and Ba, per epoch & Uniform(0,1) \\
        \hline
        \textbf{Contrast per wavelength model} & &\\
        $M$ & total mass [$M_{jup}$] & Normal(71.4, 0.6)\textsuperscript{\textdagger} \\
        $f_i, i \in 1..N_{\rm{wavelengths}}$  & contrast between Bb and Ba, per wavelength& Uniform(0,1) \\
        $C_z$  & Kernel phase spectral correlation & Uniform(0, 1) \\
        \hline
        \textbf{Uniform mass prior model} & &\\
        $M$ & total mass [$M_{jup}$] & Uniform(50, 150)  \\
        $f_i, i \in 1..N_{\rm{wavelengths}}$  & contrast between Bb and Ba, per wavelength& Uniform(0,1) \\
        $C_z$  & Spectral correlation parameter & Uniform(0, 1) \\
        \hline
    \end{tabular}
    {\raggedright \textsuperscript{*} \citet{gaiadr3}.
    \textsuperscript{\textdagger} Total mass prior from \citet{gmbrandt}. \par}
    
\end{table*}

\subsection{Outer Model Description}

\begin{table}
    \caption{Relative astrometry collated from \citet{tdbrandt_2020} and \citet{brandt_orvara_2021}. The 2024 epoch is new from this work.\label{tab:astrom}}
    \centering
    \begin{tabular}{|l|r|r|}
        \hline
        \textbf{date} (UTC) & separation (mas) & position angle ($^\circ$) \\
        \hline
        1995-11-17     & $7777.0 \pm 1.7 $ &  $163.224 \pm 0.015$ \\
        1996-05-25     & $7732.0 \pm 2.0 $ &  $163.456 \pm 0.019$ \\
        1996-11-09     & $7687.7 \pm 1.5 $ &  $163.595 \pm 0.015$ \\
        1999-05-26     & $7458.3 \pm 1.6 $ &  $164.796 \pm 0.015$ \\
        2000-05-26     & $7362.8 \pm 1.6 $ &  $165.244 \pm 0.016$ \\
        2000-11-16     & $7316.9 \pm 1.6 $ &  $165.469 \pm 0.016$ \\
        2011-03-26     & $6210.0 \pm 10.0$ &  $171.2   \pm 0.100$ \\
        2020-10-24     & $4922.1 \pm 2.3 $ &  $179.564 \pm 0.024$ \\
        2021-01-05     & $4890.5 \pm 2.4 $ &  $179.735 \pm 0.024$ \\
        2024-10-24     & $4305.6 \pm 30.0$ &  $184.03  \pm 0.4  $ \\
        \hline
    \end{tabular}
\end{table}

In order to calculate the mutual inclination, we also fit an independent outer model of the orbit of B around A, essentially reproducing the work of \citet{gmbrandt}.
We include RV data from HARPS \citep{trifonov_harps_2020}, HIRES \citep{tal-or_hires}, Lick \citep{fischer_lick_twenty_five}, and UVES \citep{uves_data}. We include the same additional HIRES RVs from \citet{cls_rvs} collated by \citet{gmbrandt}.

We split the data into sub-series with separate instrumental zero-point offsets, based on established criteria: 
J vs. H for HIRES, 
before and after the fiber upgrade for HARPS,
and split by dewar for Lick.

We include the same relative astrometry collated by \citet{gmbrandt} and \citet{tdbrandt_2020}, from Hubble Space Telescope Wide Field and Planetary Camera 2, Keck NIRC2, and Subaru HiCIAO \citep[the SEEDS survey, ][]{seeds}.
To this data, we add one relative astrometry data point from new Keck observations carried out on 2024-10-24, giving $\Delta \mathrm{R.A.} = -302.4 \pm 29$ mas and  $\Delta \mathrm{Dec.} = -4295.0 \pm 29 $ mas.
Finally, we include proper motion anomaly data from the Hipparcos-Gaia catalog of accelerations \citep[the HGCA;][]{hgca}. 
The primary goal of this effort is to ensure that we are using the same physical reference system between the inner and outer orbits, though we also slightly refine the orbit posterior based on the additional data. For reference, Octofitter uses a coordinate and convention system where the $x$ axis (R.A.) increases positively to the left (East), the $y$ axis (Dec.) increases positively upwards (North), and the $z$ axis (and radial velocity) increases positively away from the observer. This results in longitude of the ascending node measured counter-clockwise from North towards East.

In our outer orbit model, we correct for secular acceleration to both the radial velocities and proper motion anomalies by propagating the full 3D orbit of the star. We undo the static non-linearity correction applied as part of the HGCA, and apply our own dynamic correction at each epoch using the orbit and derived barycentric proper motion at each MCMC draw (referenced to J2016). We use the Gaia DR3 radial velocity of 4 km/s for this correction. For this star, these corrections have negligible impacts.

\subsection{Sampling}
Given the sparse nature of the VLTI UTs, data from any given exposure are consistent with many different companion locations and contrasts. This is especially true for shorter observing sequences where we do not benefit from the Earth rotating the interferometric array with respect to the target and filling in more of the UV-plane.

For a single epoch, it is computationally feasible to evaluate a model over a grid of 2D positions, for example to search for the peak with the highest likelihood from which to extract astrometry for subsequent orbit fitting. For data taken over longer periods of time, however, we must use an orbit model to relate a set of orbital parameters to positions in each exposure. In this case, the dimensionality of the problem is much too high for numerical integration to be feasible and we must fall back to Markov chain Monte Carlo (MCMC) methods.

Orbit models fit to relative astrometry are known to exhibit very complex posterior distributions with long tails, complex non-linear relationships between variables, and multi-modality.
Replacing astrometry (a single point at each epoch) with sparse interferometer data, which is itself highly multi-modal, multiplies this challenge significantly.

Octofitter has previously integrated the No U-Turn Hamiltonian Monte Carlo \citep[HMC;][]{hoffman_nouturn} algorithm which uses gradient information to explore posteriors very efficiently.
Unfortunately, HMC is not efficient at sampling from widely separated modes and so is not suitable for this problem \citep{hmc_multimodal_scaling}.

For highly multi-modal problems, parallel tempered sampling has emerged as a strategy for global exploration.
In parallel tempering, many chains are initialized to sample from different related distributions described by a ``temperature'' parameter.
The temperature parameter controls an interpolation between the posterior distribution and a reference distribution that is easier to explore---often chosen to be the prior. Then, the chains with high temperatures approximate the reference and low temperature distributions approximating the posterior.
By allowing samples to propose swaps between parallel chains with different temperatures, the target chain (exploring the posterior) can make jumps to other peaks in the parameter space.
One implementation of this approach that has been used by \verb|orbitize!| \citep{blunt_orbitize_2020} and \verb|orvara| \citep{brandt_orvara_2021} is 
affine-invariant ensemble sampling in \verb|ptemcee.py| \citep{foreman-mackey_emcee_2013,vousden_pt_2016}.

Recent developments in MCMC sampling research have found that the traditional tempering scheme can be sub-optimal, due to the tendency of samples to swap back and forth between adjacent temperature chains producing a diffusive, random walk effect \citep{nonreversept}. 
Instead, a better approach which we employ in this work is to use non-reversible parallel tempering \citep{nonreversept}. 
In this scheme, swaps between chains of different temperatures take place through a deterministic schedule which leads to asymptotically more samples from the prior chain arriving at the target chain. Importantly, this improvement maintains the correct stationary distribution of the tempering process.

We integrated the Pigeons \citep{pigeons} MCMC sampler into Octofitter \citep{octofitter}, which implements non-reversible tempered sampling.
Pigeons implements both gradient-based (similar to HMC) and gradient-free local exploration methods.
For this project, we used a gradient-free slice sampler \citep{neal2003slice} for local exploration due to performance challenges with automatic differentiation and certain matrix operations used in the kernel phase modelling.

In addition, we employed stabilized variational parallel tempering \citep{stabvarpartemp} which uses an additional set of parallel tempered chains between the target and a variational approximation. The variational approximation increases the rate at which samples swap to the target chain by more closely matching the target distribution compared to the prior.
Combined, these advances in MCMC sampling techniques were key enabling steps in this project, bringing sampling time from months or more to only 1-2 weeks.

\section{Results}

We now discuss our modelling results, first for the inner Ba/Bb binary, followed by models of the outer orbit around A, and finally their mutual inclination.

\subsection{Inner Orbit Modelling (Ba and Bb)}

\begin{figure*}
    \centering
    \includegraphics[width=0.353\linewidth]{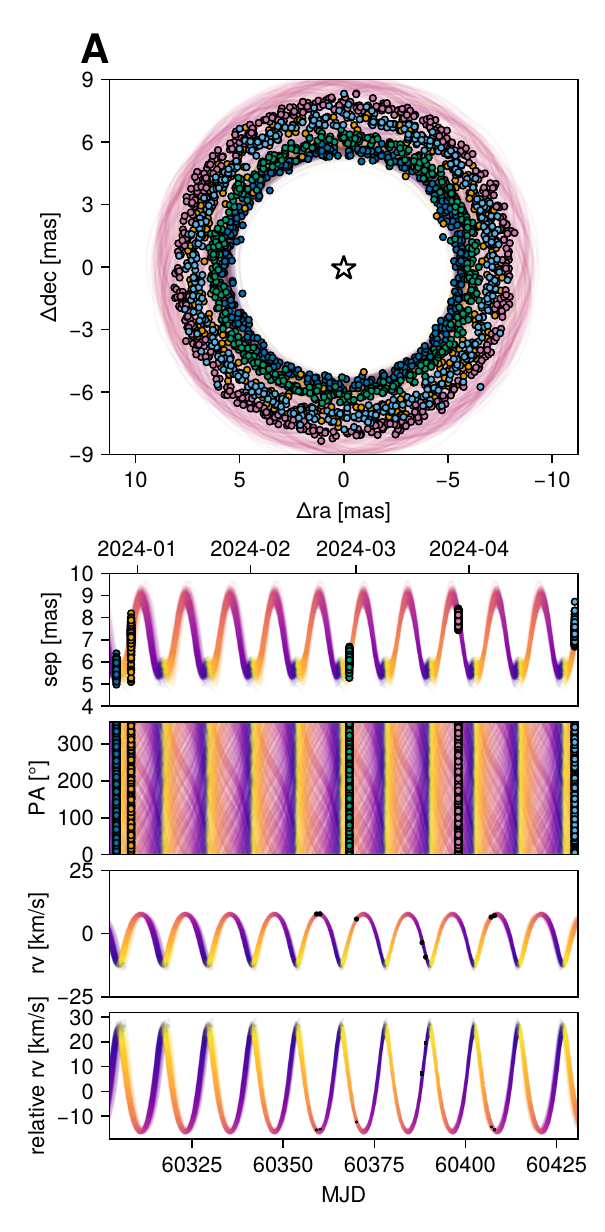}
    \hspace{-8pt}
    \includegraphics[width=0.32\linewidth]{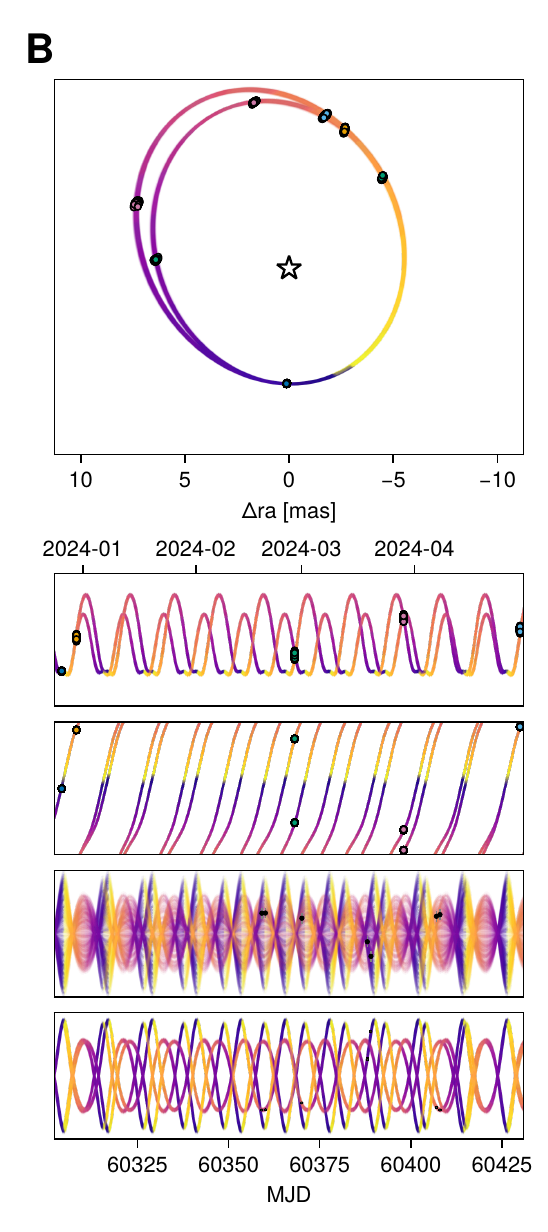}
    \hspace{-8pt}
    \includegraphics[width=0.32\linewidth]{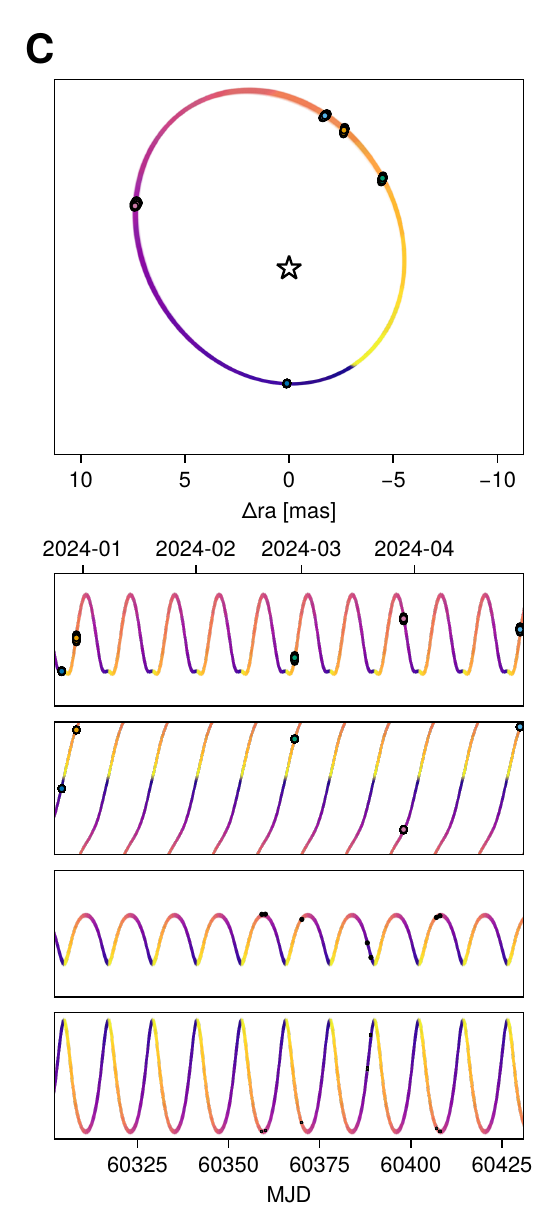} 
    \includegraphics[width=0.55\linewidth]{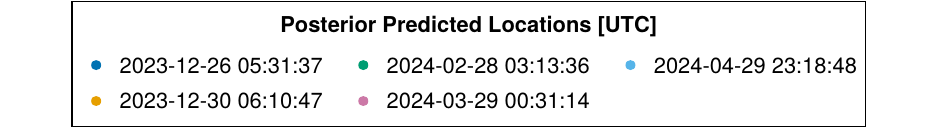}
    \includegraphics[width=0.32\linewidth]{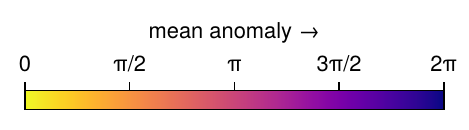}
    \caption{Comparison between constraints provided by the CRIRES+ RV data only (\textbf{A}), the GRAVITY kernel phase data only (\textbf{B}), and the combination of both (\textbf{C}). The rows present in turn the visual orbit of Bb around Ba, their projected separation, the position angle of Bb around Ba, the barycentric radial velocity of Ba, and the relative radial velocity of Bb - Ba. 
    The colour of the lines corresponds to mean anomaly, that is, the progress in time of the companion Bb in its orbit around Ba, with 0 marking periastron (closest approach) and $2\pi$ marking one full orbital period.
    The coloured dots mark the posterior values at mean times of the five GRAVITY epochs. The RV data points are plotted in black.
    With only the GRAVITY data, the posterior has four modes as seen in the bottom two RV panels.
    Part of the multimodality comes from the usual sign ambiguity in motion towards or away from the observer from astrometric data alone, and the other is a period ambiguity caused by position-angle ambiguity in the third and fourth GRAVITY epochs. The RV-only posterior has excellent agreement with one of these four modes.}
    \label{fig:compare-rv-kp}
\end{figure*}

\begin{figure*}
    \centering
    \includegraphics[width=\linewidth]{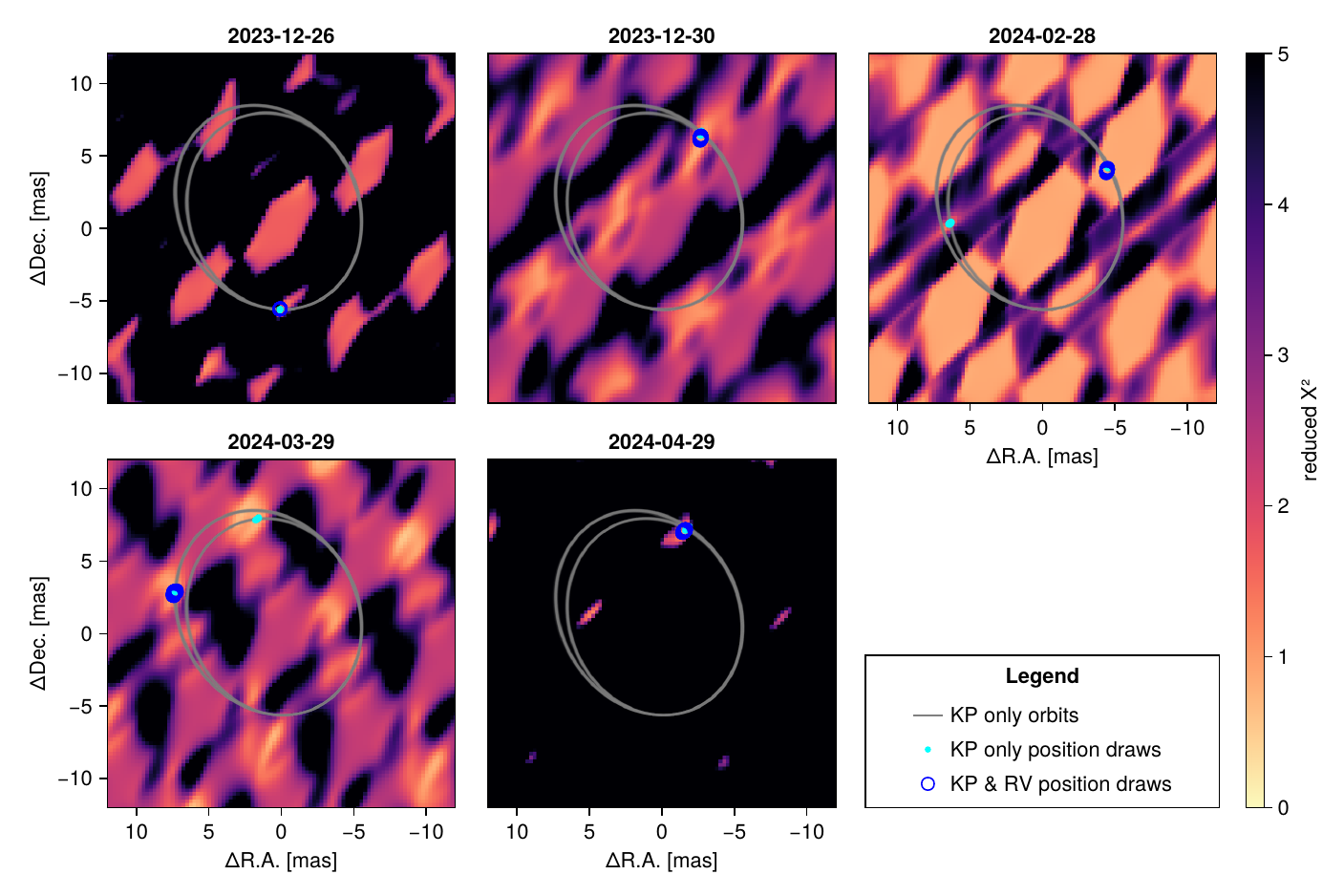}
    \caption{Kernel phase detection maps calculated at each epoch. 
    Looking at the GRAVITY data alone, there are two ranges of orbital period that are equally likely. This is because the 2024-02-28 and 2024-03-29 cannot distinguish between two different position angles, though these locations have essentially the same separation. 
    Thankfully, the addition of the radial velocity data resolves this ambiguity.
    The maps were generated by adopting the maximum \emph{a-posteriori}  spectral correlation, contrasts and kernel phase jitters per epoch.}\label{fig:detection-maps}
\end{figure*}

\begin{figure*}
    \centering
    \includegraphics[width=\linewidth]{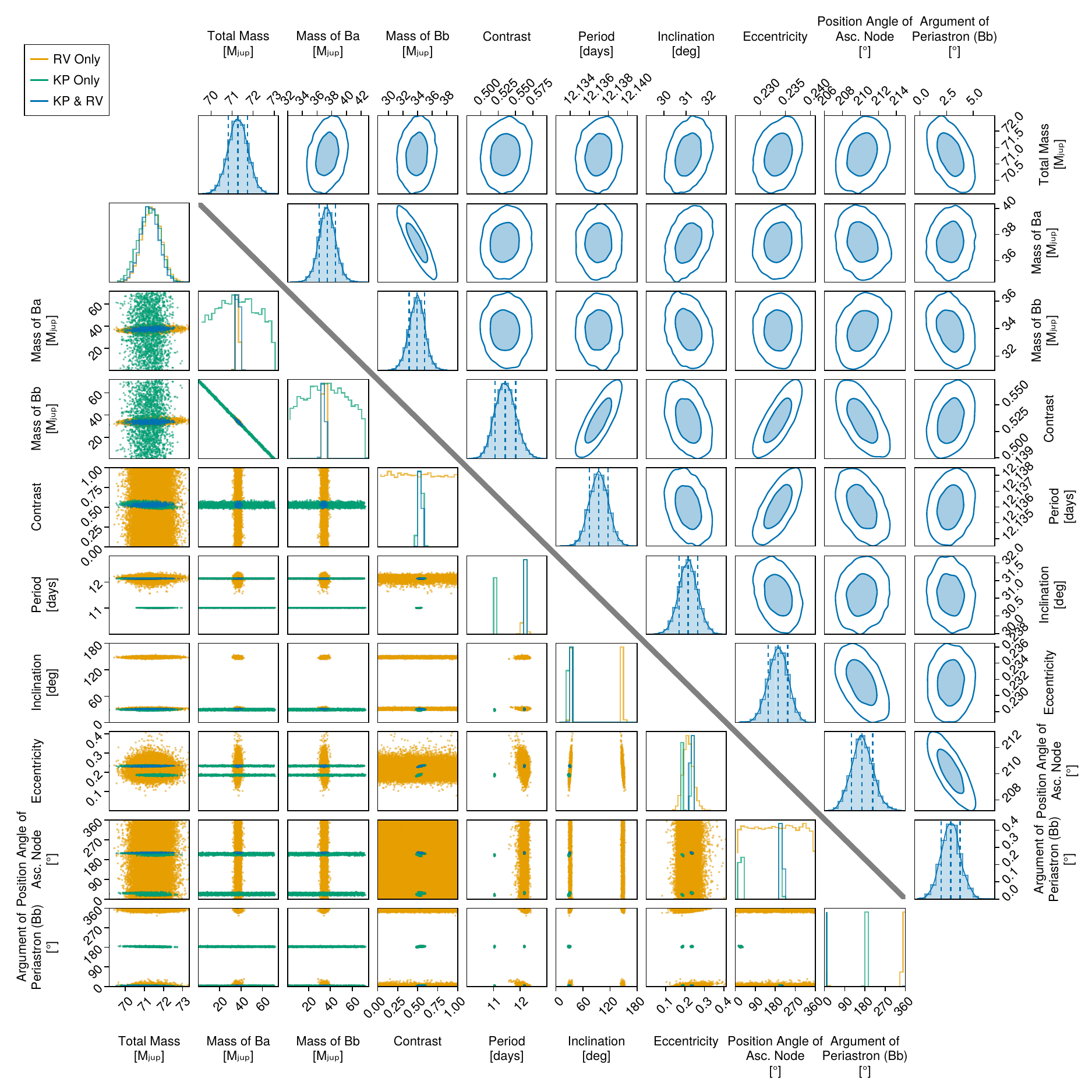}
    \caption{\textbf{Bottom left:} Corner plot comparing key variables from three different models. The model with only RV data is shown in orange, only GRAVITY kernel phase data shown in green, and both GRAVITY and RV data shown in blue. 
    Note that while RV data alone cannot constrain inclination, it can when combined with a prior on the total mass of Ba + Bb we adopted from \citet{gmbrandt}.
    The KP only posterior has four modes, one of which overlaps with one of two modes from the RV only posterior. There is very good agreement between the RV and GRAVITY posteriors, and combining them together results in the unimodal distribution shown in blue. 
    \textbf{Top right:} Same as the KP \& RV series (blue) in the bottom left with scale adjusted to show detail. These are the results originally presented in \citet{xuannat}. The contours correspond to $1\sigma$ and $3\sigma$.}
    \label{fig:kp-rv-corner}
\end{figure*}

\subsubsection{Comparing the RV, KP, and joint posteriors}
We start by assessing the agreement between the GRAVITY and CRIRES+ RV datasets, which we find are highly consistent.
Figure \ref{fig:compare-rv-kp} plots orbits sampled from the posteriors of the three different models.

Beginning with only the RV data, we find a period of $12.18\pm0.05$ days and an eccentricity of $0.215\pm0.03$.
The projected separation is constrained between 5 and 9 mas, depending on the epoch. The RV jitter parameter is consistent with $73_{-69}^{+221}$ m/s.
These results do depend strongly on the total mass prior we adopted from \citet{gmbrandt}, and other orbital periods would be plausible without it.
As we will discuss below, our other models are consistent with that prior so we do not consider its impact here.

Next, with only the KP data, we find a multi-modal posterior distribution. 
The marginal posterior is bimodal in period, consistent with either $11.032 \pm0.001$ days or $12.137\pm0.001$ days.
Figure \ref{fig:detection-maps} presents kernel phase detection maps calculated at each of the five GRAVITY epochs with orbit posteriors over-plotted.
Examining these detection maps, we note that in two of the epochs with the least visibility plane coverage, the kernel phase data are both consistent with two different locations.
The number of samples is nearly the same between the two modes of the posterior.
We also note that the two locations have essentially the same separation, but different position angles.
The KP data is furthermore bimodal in argument of periastron and in position angle of ascending node since, just like relative astrometry, these data cannot determine if the objects are receding or approaching towards the viewer.
In Figure \ref{fig:compare-rv-kp}, we calculate the posterior predictive distribution versus the RV data. Despite not being included in the model, the relative RV data can be seen by eye to match exactly one of the four posterior modes. Such chance agreement would seem unlikely, and we take this as strong confirmation that we have identified the correct orbital period.

Finally, with both the RV and KP data, we find a uni-modal posterior. The full set of orbit parameters are all constrained to near-Gaussian distributions. 
Figure \ref{fig:kp-rv-corner} compares the marginal posterior distributions from all three models. We find that orbital parameters independently constrained by each dataset agree perfectly with each other.
We draw attention particularly to the inclination versus period sub-panel and accompanying marginal histograms in that figure, where it can be seen plainly how the KP data resolve the bi-modality in the RV inclination, and the RV data resolve the bimodality in the period, leading to the uni-modal joint posterior. Note that the blue markers from the ``KP \& RV'' model perfectly overlap the green markers from one mode of the ``KP only'' posterior.

\subsubsection{Comparing GRAVITY modelling assumptions}

\begin{figure}
    \centering
    \includegraphics[width=\linewidth]{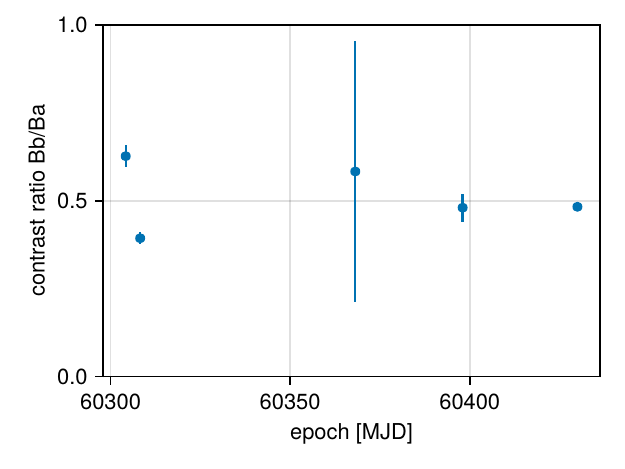}
    
    \caption{contrast between Bb and Ba versus GRAVITY epoch. In this model, the contrast is averaged over wavelengths but allowed to vary over time.}
    \label{fig:contrast-time}
\end{figure}

\begin{figure*}
    \centering
    \includegraphics[width=\linewidth]{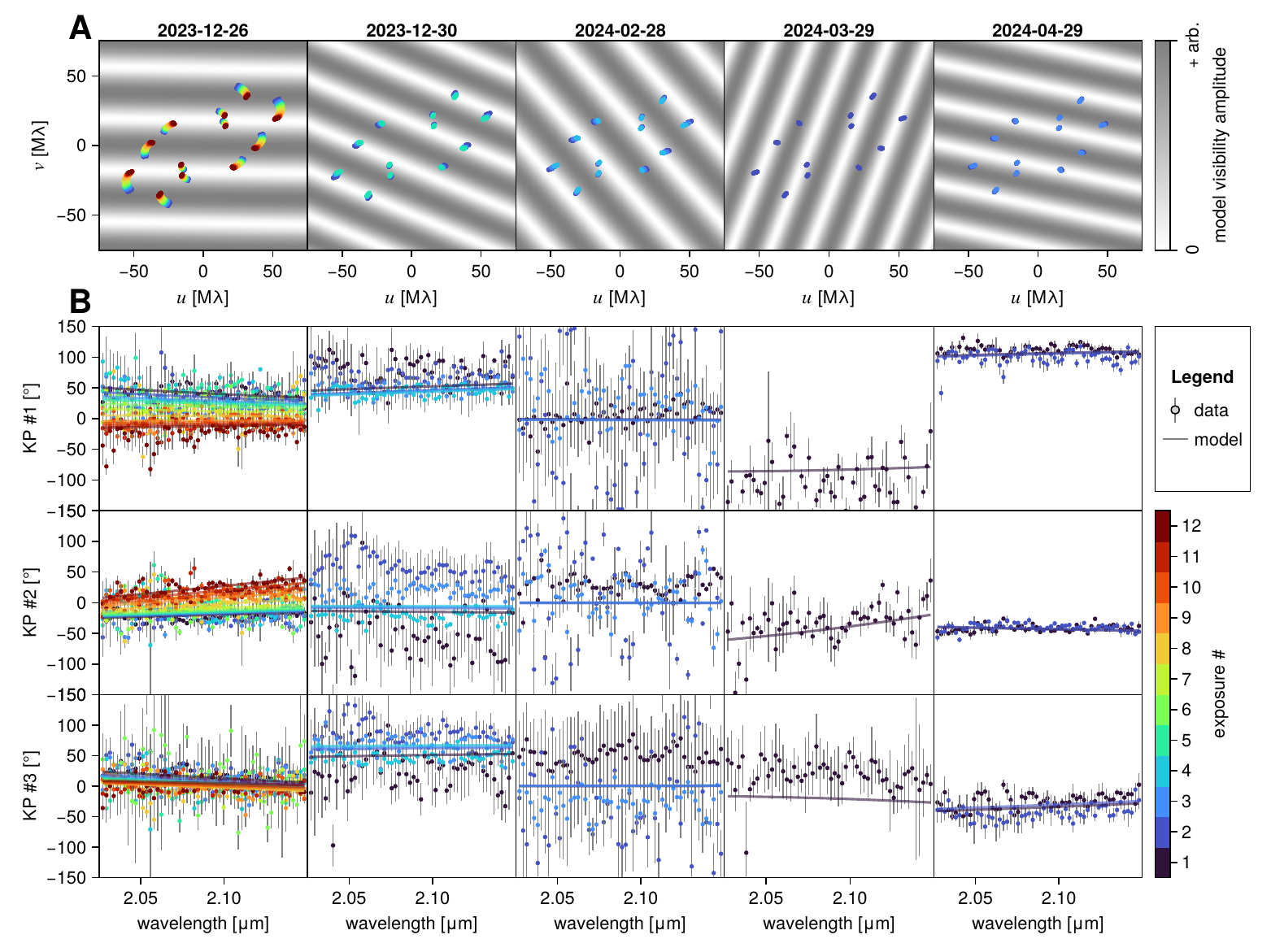}
    \caption{In-depth comparison of variable contrast model to data to aid in diagnosing the observed contrast variations. \textbf{A:} $u-v$ plane coverage at each epoch overtop the binary model visibility amplitude. \textbf{B:} Kernel phase data (points) and the variable contrast model (lines) versus wavelength, separated by epoch and kernel phase index. Notably, the first epoch, though high SNR, only marginally resolves the companion towards the second half of the epoch. This may explain why we find a higher contrast at this epoch.}
    \label{fig:vis-like-figs}
\end{figure*}

\begin{figure}
    \centering
    \includegraphics[width=\linewidth]{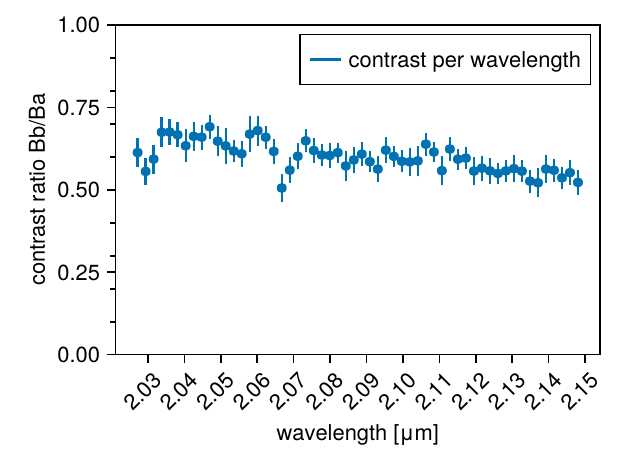}
    \caption{Marginal distribution of spectral contrast between Bb and Ba from the ``flexible contrast vs. wavelength'' model.}\label{fig:contrast-spectrum}
\end{figure}

\begin{figure*}
    \centering
    \includegraphics[width=\linewidth]{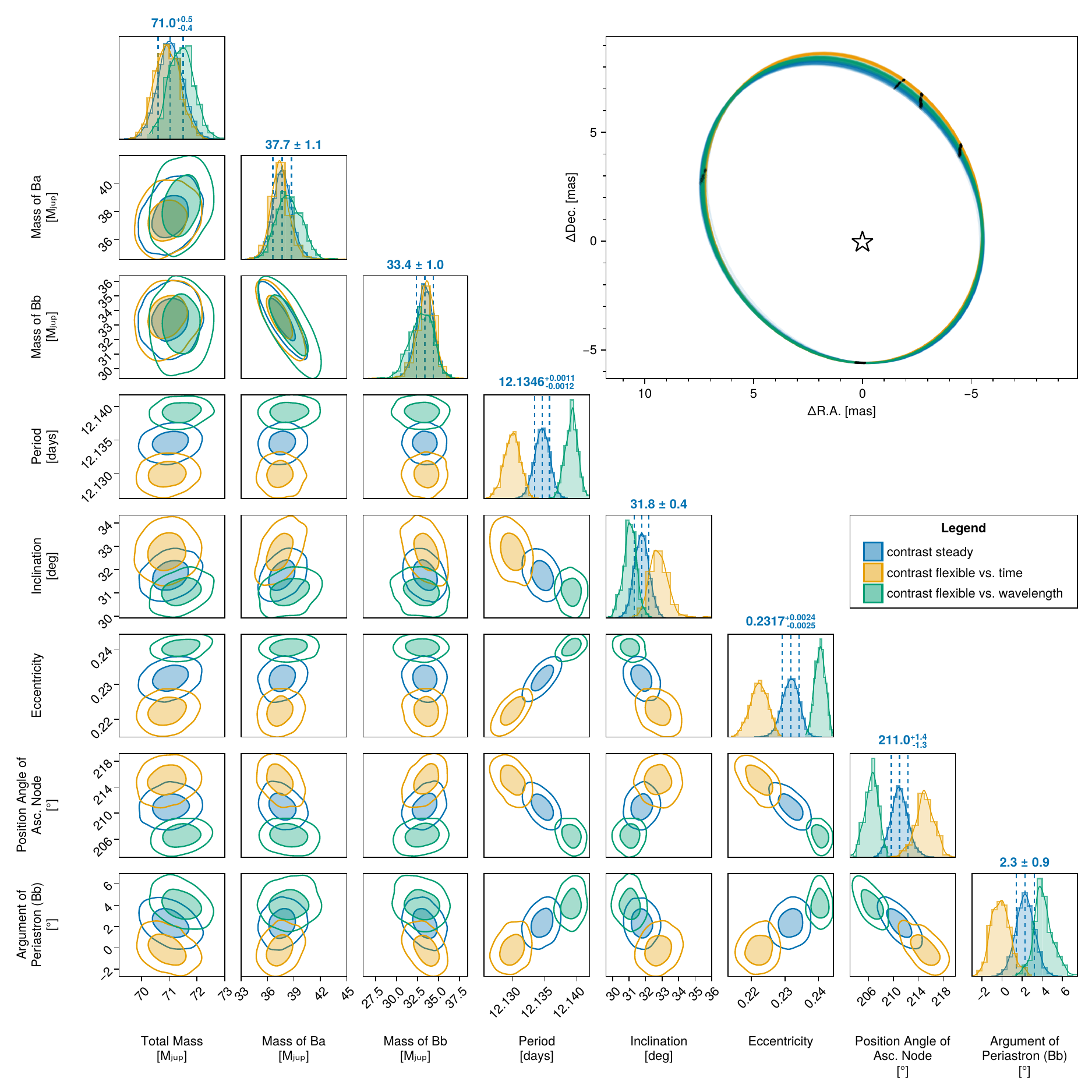}
    \caption{Comparison of orbit posteriors between the three flux models, considering variation in either spectral contrast or variation over time.}
    \label{fig:corner-cont}
\end{figure*}

In the results presented above, the contrast (that is, the flux ratio of Bb versus Ba) was fit using a single variable for all wavelengths considered and all epochs.  We now consider how contrast variations, in either time or wavelength, might affect the inferred orbital parameters.  These models also allow the spectral correlation between kernel phases, $C_z$, the kernel-phase analogue to closure-phase $C_y$ parameter proposed by \citet{jens}, to be non-zero.

The first interesting result is that in all cases $C_z$ takes on a value of $(2.3 \pm 0.3)\times10^{-3}$. 
We propose that future works could simply adopt this value for spectral correlation rather than fitting it as part of the model.

Considering now the ``contrast flexible vs. time'' model, we find a significantly higher contrast at the first epoch than the value from the ``steady contrast'' model. The contrast vs. time is plotted in Figure \ref{fig:contrast-time}. As previously discussed, we do not expect any astrophysical sources of variation would be detectable in these observations. Therefore, these variations must be caused by systematics in the interferometric data. 

To diagnose these variations, we plot the raw kernel phase data from all GRAVITY observations in Figure \ref{fig:vis-like-figs}.  Included in this panel are visualizations of the baselines in the UV plane for each exposure. We note that in 2023-12-26, the visibility amplitude of the binary is only within the longest baseline for approximately half of the exposures, meaning that it is only marginally resolved by GRAVITY. As a result, this first epoch of observations is in a regime where separation and contrast may be degenerate, which may explain the contrast discrepancy. The overall weighted mean and standard error on the mean of these per-epoch contrast values is broadly consistent with the ``steady contrast'' model.
An additional source of systematic error could be inaccurate fiber-positioning.

Considering now the ``contrast flexible vs. wavelength'' model, we find a relatively flat contrast with wavelength over the narrow range considered in our models. We do detect a significant dip in contrast at approximately 2.065 $\mathrm{\mu m}$. The posterior spectral contrast is plotted in Figure \ref{fig:contrast-spectrum}. We refrain from studying the recovered spectrum from the GRAVITY data, as future work could examine the much higher quality CRIRES+ spectra.

To compare the posteriors between the three models, we present orbits drawn from each of their posteriors in Figure \ref{fig:corner-cont}, along with a corner plot. We find that the different modelling assumptions result in similar, but not perfectly consistent orbital parameters. Notably, the separation in the 2024-04-29 epoch seems sensitive to these choices. This epoch has poor u-v coverage, but high SNR so it is not surprising that it is the most impacted. 

We further compare the Bayesian evidence of each model, estimated by Pigeons using the stepping stone approximation \citep{steppingstone}. We find a log-evidence of -17176 for the ``steady contrast'' model, -17161 for ``contrast flexible vs. time'' model, and -17272 for the ``contrast flexible vs. wavelength'' model. 
The evidence ratio strongly prefers the ``contrast flexible vs. time'' model, by a factor of $\exp((-17161) - (-17176))=3\times 10^{6}$. The ``contrast flexible vs. wavelength'' model is extremely disfavoured. 

This model comparison indicates that the contrast variability between epochs is significant, and that the data are much better fit by allowing the contrast to vary over time. This variability is almost certainly a result of uncalibrated systematics in the GRAVITY data, but the question remains--does allowing the model to adjust to these systematics produce a more, or less accurate reconstruction of the orbit?
The orbital period is sensitive to these modelling choices, so future  radial velocity or astrometric observations should be able to confirm which of these choices is the most correct.
That said, we note that allowing the contrast to vary by epoch has almost no impact on key physical parameters including the total mass and mass ratios of GL229 Ba \& Bb.
In the interest of simplicity, we quote our final posterior ranges from the simplest ``contrast steady'' model.
Future work could also study if its feasible to analytically marginalize over the contrast both per-epoch, and per-wavelength, to give the most conservative uncertainties.

Finally,  we examine the impact of the strong prior placed on the total mass of Ba + Bb, sourced from the outer orbit model \citet{gmbrandt}. Using a uniform prior on total mass, we find a credible range of consistent with $70^{+0.9}_{-0.8} \; \mathrm{M_jup}$, which agree to $1.1\sigma$. This results in slightly lower and more uncertain values for the masses of each component, and overall very similar orbital parameters with only very slight differences in $\omega$ and $\Omega$. This result in particular builds confidence that we determined the correct orbital period for the binary. 

The posterior credible ranges of the ``steady contrast'' and ``uniform mass prior'' models are both presented in Table \ref{tab:post-2}. 




\begin{table*}[]
\caption{Posterior parameter credible ranges for the inner orbit of Ba/Bb from the ``steady contrast'' model, which adopts the \citet{gmbrandt} total mass prior, and the ``uniform mass'' model, which is not informed by the outer orbit. The quoted credible intervals are the median and 0.16-0.84 quantile range.\label{tab:post}}
    \centering
        
    \begin{tabular}{|l|l|l|l|l|}
        \hline
        \textbf{Parameter}  & \textbf{Description}                      & \textbf{Credible Range}           & \textbf{Credible Range}         & \textbf{Unit} \\
                            &                                           & \textbf{Steady contrast model}    & \textbf{Uniform Mass Prior}     &               \\
        \hline
        $M$                 & total mass                                & $71^{+0.4}_{-0.5}$                & $70.0^{+0.9}_{-0.8}$            & $\mathrm{M_{jup}}$ \\
        $q$                 & mass ratio of Bb vs. Ba                       & $0.470\pm0.014$                   & $0.473 \pm 0.015$               & \\
        $M_{Ba}$            & mass of Ba                                & $37.7 \pm 1.1$                    & $36.9 \pm 1.3$                  & $\mathrm{M_{jup}}$ \\
        $M_{Bb}$            & mass of Bb                                & $33.4 \pm 1.0$                    & $33.2^{+1.0}_{-1.1}$                  & $\mathrm{M_{jup}}$ \\
        $\bar{\omega}$      & parallax                         & $173.573 \pm 0.017$               & $173.574 \pm 0.017$             & mas\\
        $e$                 & eccentricity                              & $0.2317^{+0.0024}_{-0.0025}$      & $0.2306^{+0.0025}_{-0.0026}$    & \\
        $P$                 & period                                    & $12.1346^{+0.0011}_{-0.0012}$     & $12.1343^{+0.0012}_{-0.0013}$   & d \\
        $i$                 & inclination                               & $31.8 \pm 0.4$                    & $31.4^{+0.5}_{-0.6}$            & $^\circ$ \\
        $\omega$            & argument of periapsis of Bb               & $2.3 \pm 0.9$                     & $3.2 \pm 1.1$                   & $^\circ$ \\
        $\Omega$            & position angle of ascending node               & $211.0^{+1.4}_{-1.3}$             & $210.4^{+1.5}_{-1.4}$           & $^\circ$ \\
        $\tau$              & orbit fraction at MJD 60304.17635003472   & $0.0716^{+0.0024}_{-0.0021}$      & $0.0720^{+0.0024}_{-0.0021}$    & \\
        $\rm{jitter}$       & R.V. jitter                               & $323^{+139}_{-106}$               & $355^{+151}_{-112}$             & m/s\\
        $\rm{rv}_0$         & R.V. offset                               & $427^{+197}_{-186}$               & $449^{+216}_{-196}$             & m/s\\
        $C_z$              & Kernel phase spectral correlation         & $(23.2 \pm 2.8)\times 10^{-4}$    & $(23.4 \pm 2.8)\times 10^{-4}$  & \\
        $f$               &  contrast between Bb and Ba         & $0.514^{+0.017}_{-0.015}$         & $0.511^{+0.017}_{-0.015}$       & \\
        $KP_{jit,1}$        & Kernel phase jitter, epoch 1              & $12.6 \pm 0.3$           & $12.6 \pm 0.3$                  & $^\circ \times 1.333$\\
        $KP_{jit,2}$        & Kernel phase jitter, epoch 2              & $29.8^{+1.1}_{-1.0}$              & $29.8^{+1.1}_{-1.0}$            & $^\circ \times 1.333$\\
        $KP_{jit,3}$        & Kernel phase jitter, epoch 3              & $53.6^{+2.3}_{-2.1}$              & $53.5^{+2.4}_{-2.1}$            & $^\circ \times 1.333$\\
        $KP_{jit,4}$        & Kernel phase jitter, epoch 4              & $29.8^{+2.8}_{-2.5}$              & $29.3^{+2.9}_{-2.5}$            & $^\circ \times 1.333$\\
        $KP_{jit,5}$        & Kernel phase jitter, epoch 5              & $8.7 \pm 0.6$                     & $8.7 \pm 0.6$                   & $^\circ \times 1.333$\\
        \hline
    \end{tabular}
\end{table*}

\subsection{Outer Orbit Modelling (A and B)}

\begin{figure*}
    \centering
    
    \includegraphics[width=\linewidth]{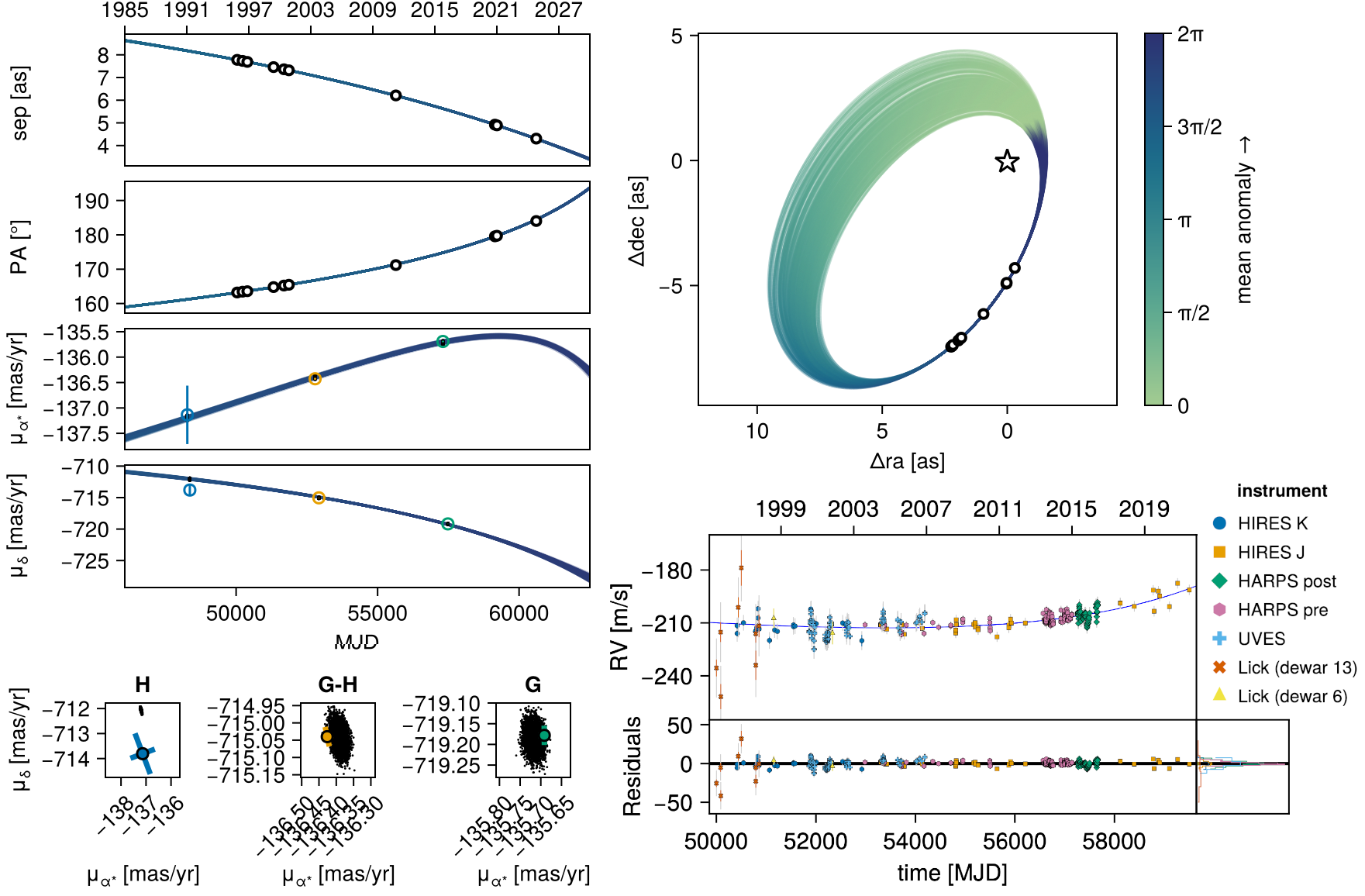}
    \caption{Outer orbit posterior of the tight binary B around the primary A updated with new relative astrometry. 
    The panels in the left column show the relative separation, position angle, and apparent proper motion induced by the binary. The bottom three sub-panels plot the residual proper motion at the Hipparcos (``H''), Gaia (``G''), and Gaia-Hipparcos (``G-H'') epochs.  The top right panel shows the projected orbit of B relative to A. The colorschemes in these plots indicate the mean anomaly of each orbit draw, progressing from 0 (periastron) to $2\pi$ (a complete orbit). 
    The bottom right panels show the radial velocity data fit of one of the posterior samples, showing an upturn in RV around 2020. The solid color errorbars indicate the data uncertainty, and the grey error bars indicate the jitter and data uncertainty added in quadrature. 
    }
    \label{fig:GL229A}
\end{figure*}

\begin{figure*}
    \centering
    \includegraphics[width=0.75\linewidth]{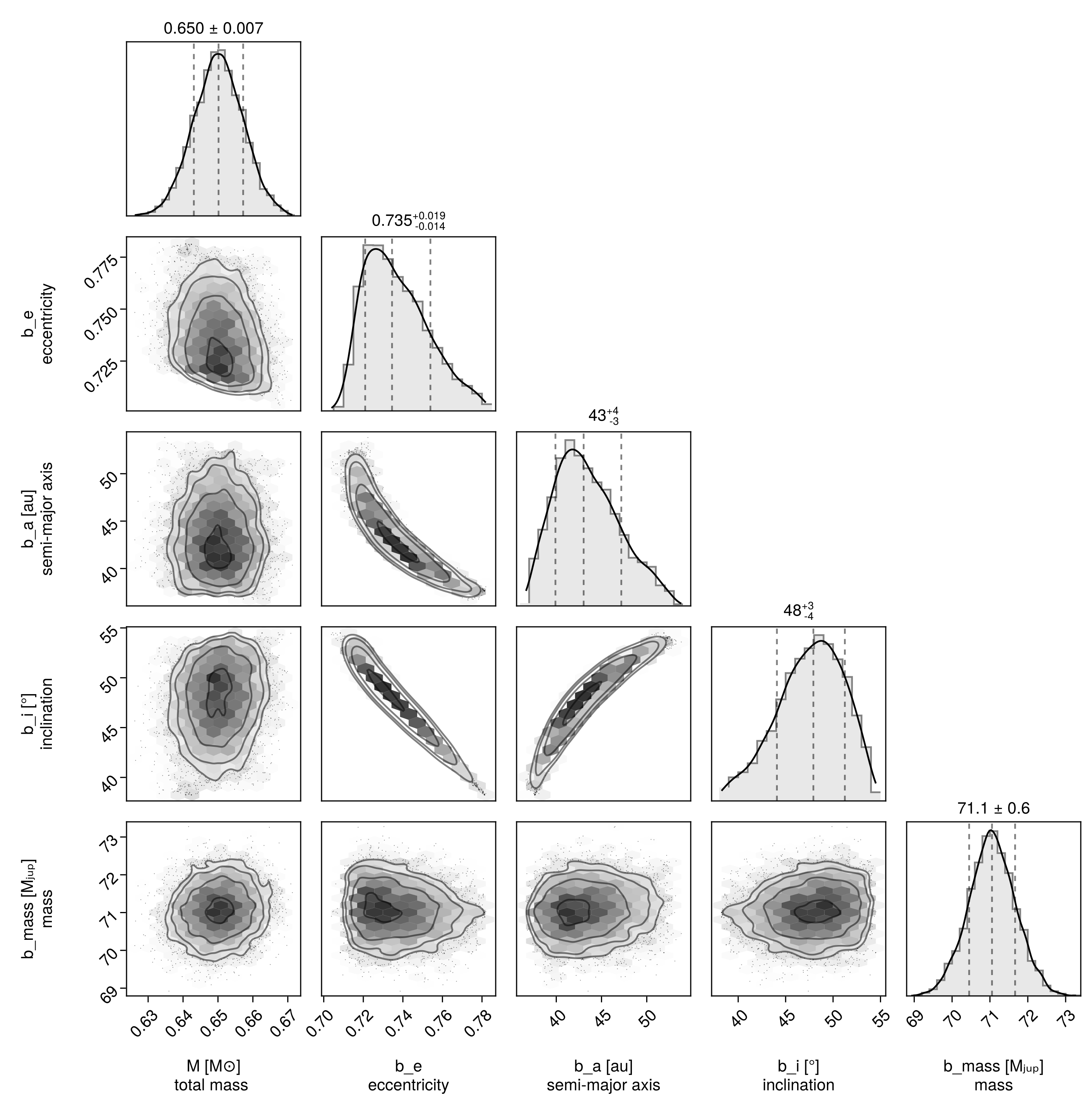}
    \caption{Corner plot of selected variables from the outer orbit posterior. Despite agreeing on the secondary mass, we find substantially higher semi-major axis and lower eccentricity than \citet{tdbrandt_2020} and \citet{gmbrandt}.}
    \label{fig:outer-corner}
\end{figure*}

The results of our outer model fit are presented in Figures \ref{fig:GL229A}, \ref{fig:outer-corner}, and Table \ref{tab:post-2}.
We find overall somewhat different results than \citet{gmbrandt}, though still consistent with the earlier work of \citet{tdbrandt_2020}. We find a higher semi-major axis ($\approx 43$ vs. $\approx 33$ AU), lower eccentricity ($\approx 0.7$ vs. $\approx 0.85$), and much higher inclination ($\approx47$ vs. $\approx 10^\circ$).
That said, we agree quite well on the total mass of the B component ($71.7\pm0.6$ vs. $71.4\pm0.6$ $\mathrm{M_{Jup}}$).

\begin{figure}
    \centering
    \includegraphics[width=\linewidth]{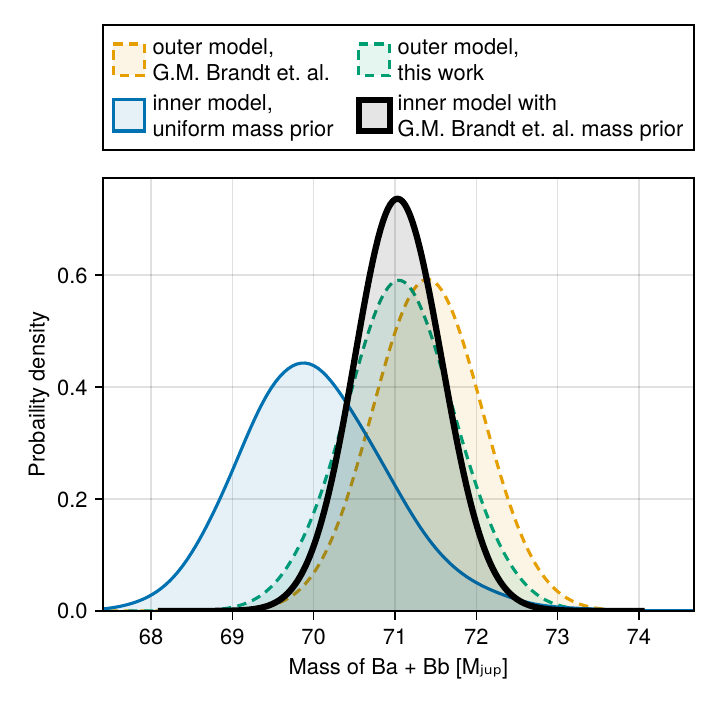}
    \caption{Mass posteriors from this work and \citet{gmbrandt}. The dashed lines show the mass as determined only by the reflex motion of the primary star A. The solid blue line shows the mass as determined only by the radial velocities of the binary Ba and Bb.
    The solid black line is the combination of these two measurements (the ``steady contrast model,'' to be precise) which updates the \citet{gmbrandt} (gold) prior from outer orbit with the new information from the inner orbit (blue).}
    \label{fig:total-mass-compare}
\end{figure}

We compare the total mass of Ba + Bb across four different models in Figure \ref{fig:total-mass-compare}.
Our outer model of B fit to relative astrometry from direct imaging, to HARPS and HIRES RVs, and to the HGCA finds a total mass for B of $71.7 \pm 0.6 \;\mathrm{M_{jup}}$ when using a standard Log-Uniform prior truncated between $1..500\;\mathrm{M_{jup}}$.
This is close to the mass posterior of $71.4 \pm 0.6 \;\mathrm{M_{jup}}$  found by \citet{gmbrandt} using essentially the same data and an unbounded Log-Uniform prior on mass.
Our inner model fit of the binary to GRAVITY and CRIRES+ data finds a total mass of $70 ^{+0.9}_{-0.8} \;\mathrm{M_{jup}}$ using uniform prior on total mass truncated to $50..150 \;\mathrm{M_{jup}}$.
Our same inner model, adopting the \citet{gmbrandt} mass posterior as a prior, is  $71 ^{+0.4}_{-0.5} \;\mathrm{M_{jup}}$.
This latter value represents our best combined estimate of the true total mass of B, using information from both the inner and outer orbits.

We highlight that despite the good agreement between the inner and outer orbit models, there is an approximately 2 mas/yr offset between the Hipparcos proper motion and posterior values, just as there were in \citet{gmbrandt}.
Besides this one data point, the rest of the data are well-modelled by the orbit of B. 
This deviation cannot be explained by three-body interactions from the known companions because the period of the Ba/Bb binary is very short compared to the measurement time-spans of Hipparcos and Gaia.
We note that such an offset on the order of 2 mas/yr appears to be common in the literature, and may point to a systematic underestimate of the Hipparcos proper motion uncertainty in the HGCA.

\begin{table*}[]
\caption{Posterior parameter credible ranges for the outer orbit model. The quoted credible intervals are the median, and 0.16-0.84 quantile range. The R.V. zero-points are marginalized out analytically and are so are not parameters in the MCMC.\label{tab:post-2}}
    \centering
    \begin{tabular}{|l|l|l|l|}
        \hline
        \textbf{Parameter}  & \textbf{Description}           & \textbf{Credible Range}                    & \textbf{Unit} \\
                            &                                & \textbf{outer orbit model}                 &               \\
        \hline
        $M$                 & total mass                     & $0.650 \pm 0.007$                          & $\mathrm{M_{\odot}}$ \\
        $M_{A}$             & mass of A                      & $0.581 \pm 0.007$                          & $\mathrm{M_{\odot}}$ \\
        $M_{B}$             & mass of B                      & $71.7 \pm 0.6$                             & $\mathrm{M_{jup}}$ \\
        $\bar{\omega}$      & parallax distance              & $173.574 \pm 0.017$                        & mas\\
        $a$                 & semi-major axis                & $42.9^{+3.0}_{-2.4}$                       & au \\
        $e$                 & eccentricity                   & $0.736^{+0.014}_{-0.011}$                  & \\
        $P$                 & period                         & $349^{+37}_{-29}$                          & Julian years \\
        $i$                 & inclination                    & $47.7^{+2.5}_{-2.9}$                       & $^\circ$ \\
        $\omega$            & argument of periapsis of B     & $132 \pm 5$                                & $^\circ$ \\
        $\Omega$            & position angle of ascending node    & $168.3^{+0.3}_{-0.4}$                      & $^\circ$ \\
        $\theta$            & position angle at MJD 50000.0  & $163.153 \pm 0.007$                        & $^\circ$ \\
        $j_1$               & Jitter HIRES\_k                & $4.6 \pm 0.6$                              & m/s \\
        $j_2$               & Jitter HIRES\_j                & $2.92^{+0.31}_{-0.26}$                     & m/s \\
        $j_3$               & Jitter HARPSpost               & $3.16 \pm 0.13$                            & m/s \\
        $j_4$               & Jitter HARPSpre                & $1.97 \pm 0.07$                            & m/s \\
        $j_5$               & Jitter UVES                    & $4.8 \pm 0.3$                              & m/s \\
        $j_6$               & Jitter Lick (dewar 13)         & $21^{+5}_{-4}$                             & m/s \\
        $j_7$               & Jitter Lick (dewar 6)          & $1^{+4}_{-1}$                              & m/s \\
        $\mu_{\alpha^*}$    & barycentric proper motion in R.A. (J2016) & $-145.45 \pm 0.13$              & mas \\
        $\mu_{\delta}$      & barycentric proper motion in Dec. (J2016) & $-705.83 \pm 0.15$              & mas \\
        \hline
    \end{tabular}
\end{table*}

\subsection{Mutual inclination}

\begin{figure*}
    \centering
    \includegraphics[width=0.75\linewidth]{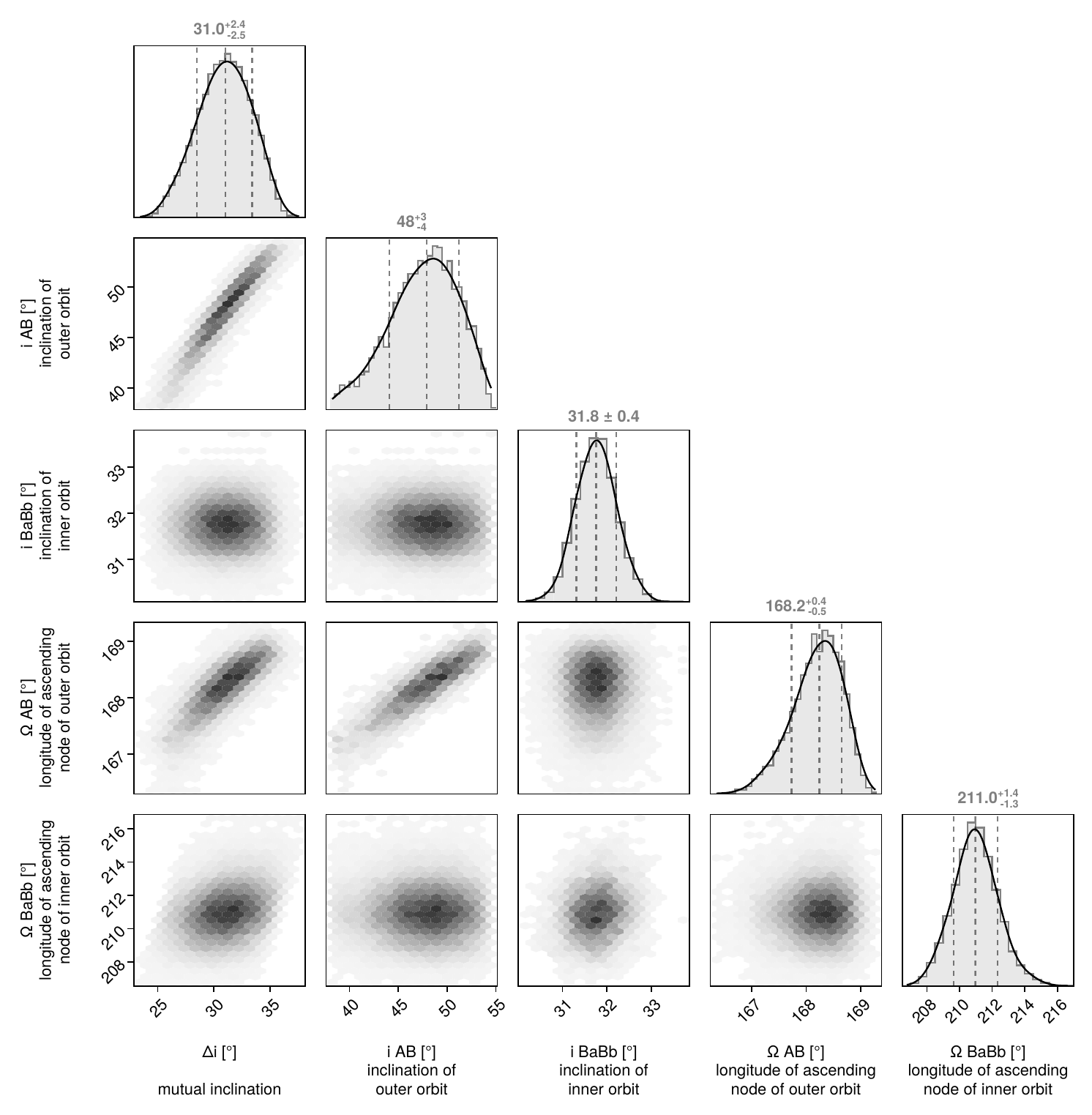}
    \caption{Mutual inclination between the inner (``steady contrast'' model) and outer orbits, and comparison of orbit orientations between the systems.}
    \label{fig:mutual-inclination}
\end{figure*}

Using the orbital parameters we inferred from our inner and outer orbits (``steady contrast'' model), we calculate the mutual inclination of the inner binary orbit compared to the outer orbit around A.
We find a mutual inclination of $31 \pm 2.5$ $^\circ$, below the $\approx39.2^\circ$ threshold for Kozai-Lidov oscillations to become significant. The mutual inclination is plotted against the inclination and position angle of ascending node of the inner and outer models in Figure \ref{fig:mutual-inclination}. As this figure shows, the uncertainty on the mutual inclination is dominated by the degeneracy between inclination and eccentricity in the outer orbit. Further orbital monitoring of the outer orbit with relative astrometry and RV would improve these constraints.

\section{Discussion}
In this work, we presented our modelling of the binary brown dwarf companion GL229 Ba and Bb from \citet{xuannat} in additional detail.
We develop and deploy new methodology to model GRAVITY closure phase data using kernel phases in order to improve the robustness of our Bayesian models.
In doing so, we show how we can jointly model orbits and raw GRAVITY data in order to constrain the orbit and contrast of an object.

We integrate the posterior globally using a novel application of non-reversible parallel tempered MCMC, making it so that we can constrain an object's orbit using multiple epochs of data, without having to select a single peak in each detection map \emph{a-priori}.
This combination of modelling an orbit directly from interferometer data, with a MCMC method that can feasibly handle the global multi-peaked posterior, reveals a connection between visibility-plane coverage and orbital phase coverage.
Future observation campaigns of fast-orbiting objects should consider using much shorter observing sequences, repeated many times over the course of an orbit, rather than trying to gather sufficient visibility plane coverage in a single night.

Our analysis reveals that the CRIRES+ and GRAVITY data are in excellent agreement. Both of these datasets produce bi-modal posteriors (seen in Figure \ref{fig:kp-rv-corner}), but combining the two into a joint posterior leaves only a single highly constrained mode. These results are also broadly consistent with the mass vs. period curve found by \citet{Whitebook_2024} at a $2\sigma$ level.
At this stage, our results now conclusively rule out the companion mass of $60.42 ^{+2.34}_{-2.38}\;\mathrm{M_{jup}}$ presented by \citet{Feng_2022}, calculated from the Gaia-Hipparcos proper motion anomaly using a different technique than \citet{tdbrandt_2020}, \citet{gmbrandt}, and the outer model presented here.

Examining our contrast modelling assumptions, we find that there is significant variation in contrast between certain epochs. While it is possible this variation is astrophysical in nature, we rather suspect that it may be a result of uncalibrated systematic errors in the data -- particularly with the first epoch, where the GRAVITY data only marginally resolves the pair at less than 5 mas separation. Fiber positioning errors may also contribute, though not at a level sufficient to explain the higher contrast  in the first epoch. Our posteriors adopt non-zero kernel phase jitters indicating the uncertainties in the GRAVITY data are underestimated. Despite these limitations, the orbit is nonetheless still very well constrained and in reasonable agreement between models with different contrast treatments. While the relative disagreements exceed approximately $1\sigma$, in absolute terms key physical parameters like mass, period, and eccentricity, and inclination are consistent to the 0.5 $\mathrm{M_{jup}}$, 0.01 days, 0.02, and $2^\circ$ levels respectively. Future work could aim to find the source of these systematic errors, or improve their calibration.

In addition to the orbit of binary GL229 Ba and Bb, we also update the outer orbit of GL229 B with a new astrometry point from Keck/NIRC2 and more archival RV data. 
We find somewhat different orbit posterior than \citet{gmbrandt}, though a very similar mass.
Our new results point to an orbit with semi-major axis of $43^{+4}_{-3}$ au, lower eccentricity of $0.735^{+0.019}_{-0.014}$, and higher inclination of $48^{+3}_{-4}\,^\circ$. Overall we find greater uncertainty in the orbit parameters than previously reported, despite the additional data.

We calculate the total mass of the binary Ba and Bb using both the inner orbit (GRAVITY and CRIRES+) and outer orbit (relative astrometry, HGCA proper motion anomaly, HARPS, HIRES, Lick, and UVES RVs), and find good, but not perfect agreement. The inner orbit, with uniform priors on the total mass, is consistent with $70^{+0.9}_{-0.8}$, while the total mass from our updated outer orbit is $71.7 \pm 0.6$. Adopting a mass value of $71.4\pm0.6$ from \citet{gmbrandt} as a prior (outer orbit), our updated posterior incorporating the inner orbit information is $71^{+0.4}_{-0.5}$.
It would be interesting for future work to use this precisely characterized system to test our ability to measure dynamical masses from RV and proper motion anomaly.

Finally, we calculate the mutual inclination from our updated inner and outer orbit models. We find a mutual inclination of $31\pm2.5$ with uncertainty dominated by the uncertainty in the outer orbit. Continued RV modelling into the 2030s would be beneficial in constraining this value.

Based on this work, we recommend that future interferometric observations of fast-orbiting systems could consider using shorter observing sequences without much regard for visibility plane coverage per-epoch, and instead use orbital motion over several epochs to constrain the orbit. 

\vspace{2em}
\textbf{Acknowledgements}

The authors thank Lindy Blackburn for helpful suggestions surrounding kernel phases.

This paper is based on observations collected at the European Southern Observatory under ESO programmes 0112.C-2369(A), 0112.C-2369(B) and 2112.D-5036(A).

Some of the data presented herein were obtained at Keck Observatory, which is a private 501(c)3 non-profit organization operated as a scientific partnership among the California Institute of Technology, the University of California, and the National Aeronautics and Space Administration. The Observatory was made possible by the generous financial support of the W. M. Keck Foundation.

The Pigeons package development and collaboration is supported by a Canadian Statistical Sciences Institute Collaborative Research Teams Grant.

This research was enabled in part by support provided by Cedar hosted at Simon Fraser University (www.sfu.ca) and the Digital Research Alliance of Canada (alliancecan.ca).

The authors wish to recognize and acknowledge the very significant cultural role and reverence that the summit of Maunakea has always had within the Native Hawaiian community. We are most fortunate to have the opportunity to conduct observations from this mountain.

\vspace{5mm}
\facilities{VLTI (GRAVITY), Keck:II (NIRC2), VLT:Melipal (CRIRES+), ESO:3.6m (HARPS),  Keck:I (HIRES),  VLT:Kueyen (UVES), HST (WFPC2), Subaru.}

\software{
Julia \citep{julialang},
Pigeons.jl \citep{pigeons},
PairPlots.jl \citep{pairplots},
Makie \citep{makie}.
}

\bibliography{ms}{}
\bibliographystyle{aasjournal}

\end{document}